\documentstyle[12pt,amsfonts,axodraw]{article}
\begin{document}
\title{\bf Perturbative BF--Yang--Mills theory on noncommutative  
${\mathbb{R}}^4$}
\author{ H. B. Benaoum  \\  
Institut f\"ur Physik , Theoretische Elementarteilchenphysik, \\      
Johannes Gutenberg--Universit\"at, 55099 Mainz, Germany \\
email : benaoum@thep.physik.uni-mainz.de
} 
\date{ }
\maketitle
~\\
\abstract{A $U(1)$ BF--Yang--Mills theory on noncommutative ${\mathbb{R}}^4$ 
is presented and in this formulation the $U(1)$ Yang--Mills theory 
on noncommutative ${\mathbb{R}}^4$ is seen 
as a deformation of the pure BF theory. Quantization using BRST symmetry 
formalism is discussed and Feynman rules are given. Computations at 
one--loop order have been performed and their renormalization studied. 
It is shown that the $U(1)$ BFYM on noncommutative ${\mathbb{R}}^4$ is 
asymptotically free and its UV--behaviour in the computation of the 
$\beta$--function is like the usual $SU(N)$ commutative BFYM and 
Yang Mills theories.    
}
~
~\\
{\bf MZ--TH/99--48 } 
~\\
\begin{center}
{\em To the memory of my father }
\end{center}
\newpage
~\\
\section{Introduction :}
It is generally believed that our picture of the space--time as a 
manifold locally viewed as a flat Minkowski space should be 
modified drastically at the Planck scale~\cite{dop}. One possibility 
is to consider that the space, at this scale, is a noncommutative 
space and field theories on these spaces have to be formulated in the 
framework of noncommutative geometry~\cite{con}. Constructions of the 
fundamental interactions along Connes's approach are described in the 
literature~\cite{cha}. \\ 
Moreover, it has been shown by Connes, Douglas and Schwarz~\cite{dou} that 
a supersymmetric Yang--Mills theory on noncommutative torus is naturally 
related to compactification of Matrix theory which means that 
noncommutative geometry can be usefull in string theory~\cite{sei}. \\
Yet, the most urgent problem waiting to be solved, is whether or not 
quantum field theory on noncommutative space is well--defined. \\
Specifically, it has been established in the case of noncommutative 
space--time, that two dimensional theories on the Fuzzy sphere~\cite{gro} 
and on the quantum cylinder possess no UV--divergences at all. In 
contrast, field theories on noncommutative ${\mathbb{R}}^4$~\cite{dem},    
noncommutative 3--tori~\cite{bon} and quantum plane have 
UV--divergences~\cite{fil}. Thus, ultraviolet behaviour of a field theory on 
noncommutative space is sensitive to the topology. \\
Recently, one--loop renormalizability of Yang--Mills on noncommutative 
${\mathbb{R}}^4$~\cite{mar} and on noncommutative torus~\cite{she} has been 
demonstrated. Particularly, it has also been shown that these fields 
theories have an asymptotical free behaviour.  
A general discussion about renormalizability of a massive scalar 
quantum field theoy on noncommutative ${\mathbb{R}}^d$ has been carried 
out in~\cite{che} and unfamiliar mixing effects of IR and UV due to 
non--commutativity has been found in~\cite{shir}. Such a mixing has no 
analog in conventional quantum field theory. \\
~\\
Here, we will be concerned with the perturbative quantization of the   
so--called BF--Yang--Mills theory ( BFYM ) on noncommtative ${\mathbb{R}}^4$. \\
We recall that BFYM formulation on commutative ${\mathbb{R}}^4$ has been used 
in~\cite{fuc} to introduce an explicit representation of the 't Hooft 
algebra, making closer connection between Yang--Mills theory and topological 
field theories of BF type~\cite{hor}. \\
The euclidean BFYM on commutative ${\mathbb{R}}^4$ is described by the action :
\begin{eqnarray}
S^{com}_{BFYM} & = & \int Tr \left( i B \wedge F + g^2~B \wedge \ast B 
\right) \nonumber \\
& = & \int d^4 x \left( \frac{i}{2}~\epsilon^{\mu \nu \alpha \beta} 
B^a_{\mu \nu} F^a_{\alpha \beta} + g^2~B^a_{\mu \nu} B^{a~\mu \nu} 
\right) .
\end{eqnarray}
where $F = F^a_{\mu \nu} dx^{\mu} \wedge dx^{\nu}~T^a$ is the usual 
field strength, $B$ is a Lie valued 2--form  and $\ast$ is the Hodge 
product for a $p$--from with $T^a$ as generators of $SU(N)$ Lie algebra in the 
fundamental representation normalized as $Tr ( T^a T^b ) = \frac{1}{2} 
\delta_{a b}$. \\
In this framework Yang--Mills theory is seen in the first order formalism 
as a deformation of a topological theory of BF type. A proof of 
the full equivalence of BFYM with the standard second order formalism 
has been achieved through path integral and algebraic methods~\cite{tan}. \\
Its UV--behaviour and computation of the $\beta$--function have been 
carried out showing that it is equal to the Yang--Mills case~\cite{zen}. \\
~\\
The paper is organized as follows. In section 2 we introduce the $U(1)$   
BFYM on noncommutative ${\mathbb{R}}^4$ and study its quantization with the 
BRST formalism. In section 3 we derive the Feynman rules and compute the 
one--loop divergent diagrams. Section 4 is concerned with the one--loop 
renormalizability and explicit one--loop computation of the $\beta$--
function. Finally in section 5, we summarize our results.
\section{BF Yang Mills on noncommutative ${\mathbb{R}}^4$ :}
The noncommutative ${\mathbb{R}}^4$ is defined as the algebra ${\cal A}_\theta$ 
generated by $x_{\mu}, \mu = 1,2,3,4$ satisfying the commutation relations : 
\begin{eqnarray}
\left[ x_{\mu},  x_{\nu} \right]~~=~~i~\theta_{\mu \nu},~~
\left[ x_{\mu},  \theta_{\rho \nu} \right]~~=~~0. 
\end{eqnarray}
where $\theta_{\mu \nu}$ is a real constant antisymmetric 
matrix with rank 4. \\ 
In the non--commutative geometry framework, the geometrical features 
of the noncommutative manifold are described by a $C^{\star}$--algebra. 
The algebra of functions on noncommutative ${\mathbb{R}}^4$ can be 
considered as the deformation of the $C^{\star}$--algebra of 
continuous complex functions over ${\mathbb{R}}^4$ vanishing at 
infinity, using the Weyl product, 
\begin{eqnarray}
( f \ast g ) ( x ) & = & e^{\frac{i}{2}~\theta_{\mu \nu} 
\frac{\partial}{\partial \xi_{\mu}} \frac{\partial}{\partial \zeta_{\nu}}}~
f( x + \xi ) g( x + \zeta) |_{\xi = \zeta = 0} \nonumber \\ 
& = & \int \int \frac{d^4 p}{(2 \pi)^2}~\frac{d^4 q}{(2 \pi)^2}~
e^{i \omega (p,q)}~e^{i ( p + q) x} \tilde{f} (p)~\tilde{g} ( q) .
\end{eqnarray} 
where $\omega ( p, q) = \frac{1}{2} \theta_{\mu \nu} p^{\mu} q^{\nu}$ 
and 
$\tilde{f}$ ( resp. $\tilde{g}$ ) is the Fourier transform of $f$ 
( resp. $g$ ) . \\ 
The $\star$ product satisfies the following identities : 
\begin{eqnarray}
\int d^4 x~( f \ast g ) ( x ) & = & \int d^4 x ( g \ast f ) ( x)~=~
\int d^4 x f(x) g(x)   \nonumber \\
\int d^4 x~( f \ast g \ast h ) ( x) & = & \int d^4 x~(h \ast f \ast g) (x)~
=~\int d^4 x ( g \ast h \ast f ) ( x) 
\end{eqnarray}
The last identity follows from the associativity of the $\ast$ product , i.e 
$( f \ast g) \ast h = f \ast (g \ast h)$. \\ 
~\\
An alternative reformulation of the $U(1)$ Yang--Mills theory on noncommutative 
${\mathbb{R}}^4$ is through the first order formalism. This model, named   
"gaussian" $U(1)$ BF--Yang--Mills on noncommutative ${\mathbb{R}}^4$  
is described by the action : 
\begin{eqnarray}
S_{BFYM} & = & \int d^4 x~\left( \frac{i}{2}~ \epsilon^{\mu \nu \alpha \beta} 
B_{\mu \nu} \ast F_{\alpha \beta} + g^2 B_{\mu \nu} \ast B^{\mu \nu} 
\right) ( x ). 
\end{eqnarray}
Here $F_{\mu \nu} = \partial_{\mu} A_{\nu} - \partial_{\nu} A_{\mu} + 
\{ A_{\mu}, A_{\nu} \}$ is the field strength for the antihermitian 
gauge field $A_{\mu}$ where $\{ A_{\mu}, A_{\nu} \}(x)  
= \left( A_{\mu} \ast A_{\nu} \right) (x) - 
\left( A_{\nu} \ast A_{\mu} \right) (x)$ is the Moyal bracket and 
$B_{\mu \nu}$ is an antisymmetric tensor. \\
It is easy to see that this action is on--shell equivalent to the classical 
$U(1)$ Yang--Mills action on noncommutative ${\mathbb{R}}^4$. This latter  
can also be recovered by performing a path integration over $B_{\mu \nu}$.  
Indeed, after a field redefinition $B \rightarrow B/g$ we get :
\begin{eqnarray}
\int {\cal D}B_{\mu \nu}~ e^{- S_{BFYM} }~ \propto ~e^{- S_{YM} }. 
\end{eqnarray}
where $S_{YM}$ is the Yang--Mills action on noncommutative 
${\mathbb{R}}^4$ given by : 
\begin{eqnarray}
S_{YM} & = & \frac{1}{4 g^2} \int d^4 x~ \left( F^{\mu \nu} \ast 
F_{\mu \nu} \right) (x). 
\end{eqnarray} 
The BFYM action is invariant under the usual gauge symmetry : 
\begin{eqnarray}
\delta A_{\mu}~~=~~D_{\mu} \epsilon~~,~~~~
\delta F_{\mu \nu}~~=~~\{ F_{\mu \nu}, \epsilon \}~~,~~~~
\delta B_{\mu \nu}~~=~~\{ B_{\mu \nu}, \epsilon \}.
\end{eqnarray}
with $D_{\mu} \epsilon = \partial_{\mu} \epsilon + 
\{ A_{\mu}, \epsilon \}$. \\ 
~\\ 
In the following we will use the BRST formalism~\cite{pig}, which requires   
the introduction of Faddeev--Popov ghost $c$ and anti--ghost $\bar{c}$, 
auxiliary field $b$ and a $s$--BRST operator defined as :
\begin{eqnarray}
s A_{\mu} & = & D_{\mu} c~,~~s F_{\mu \nu}~~=~~\{ F_{\mu \nu}, c \}~,~~
s c~~=~~ - c \ast c~,~~s \bar{c}~~=~~b~, \nonumber \\
s b & = & 0~,~~
s B_{\mu \nu}~~=~~\{ B_{\mu \nu}, c \}.
\end{eqnarray}
which is off--shell nilpotent $s^2 = 0$. \\ 
Correspondingly, one introduces a gauge--fixing term action :
\begin{eqnarray}
S_{gf} & = & \int d^4 x~ s \left( \bar{c} \ast ( \frac{\alpha}{2} b - 
\partial_{\mu} A^{\mu} ) ~\right) (x). 
\end{eqnarray} 
and an external field contribution :
\begin{eqnarray}
S_{ext} & = & \int d^4 x~ \left(~\Omega^{\mu}_A \ast s A_{\mu} + 
\Omega^{\mu \nu}_B \ast s B_{\mu \nu} + \Omega_c \ast s c 
  ~\right) (x). 
\end{eqnarray}
Then the complete tree--level action is :
\begin{eqnarray} 
\Sigma \left[ A_{\mu}, B_{\mu \nu}, c, \bar{c}, b, \Omega_{A \mu}, 
\Omega_{B \mu \nu}, \Omega_c \right] 
~=~ S_{BFYM} + S_{gf} + S_{ext}  \nonumber \\ 
~=~ \int d^4 x~ \left( \frac{i}{2}~ \epsilon^{\mu \nu \alpha \beta} 
B_{\mu \nu} \ast F_{\alpha \beta} + g^2 B_{\mu \nu} \ast B^{\mu \nu} 
\right. 
\nonumber \\
\left. + b \ast ( \frac{\alpha}{2} b - \partial_{\mu} A^{\mu} ) + 
\bar{c} \ast \partial_{\mu} D^{\mu} c 
+  \Omega^{\mu}_A \ast D_{\mu} c + 
\Omega^{\mu \nu}_B \ast \{  B_{\mu \nu}, c \} - 
\Omega_c \ast ( c \ast c ) ~\right).
\end{eqnarray}
which satisfies the Slavnov--Taylor identity : 
\begin{eqnarray}
{\cal S} \left( \Sigma \right) & = & 0. 
\end{eqnarray}
where 
\begin{eqnarray}
{\cal S} \left( \Sigma \right) & = & \int d^4 x~ \left( 
\frac{\delta \Sigma}{\delta \Omega^{\mu}_A}~
\frac{\delta \Sigma}{\delta A_{\mu}} + 
\frac{\delta \Sigma}{\delta \Omega^{\mu \nu}_B}~
\frac{\delta \Sigma}{\delta B_{\mu \nu}} + 
b~\frac{\delta \Sigma}{\delta \bar{c}} + 
\frac{\delta \Sigma}{\delta \Omega_c}~
\frac{\delta \Sigma}{\delta c} \right).  
\end{eqnarray}
\section{ One--loop calculations :}
To check the one--loop UV--behaviour of the BFYM on noncommutative 
${\mathbb{R}}^4$, we 
derive the Feynman rules by expanding $S_{BFYM} + S_{gf}$ and separating it 
into quadratic and higher order pieces. After field rescaling,   
$B \rightarrow B/g$ and $A \rightarrow g A$, the quadratic pieces give the 
propagators in momentum space for the fields $A, B$ and $c$, and the 
higher order pieces give the vertices $BAA$ and $\bar{c} A c$. \\ 
After expansion, the quadratic action is found to be :    
\begin{eqnarray}
S_0~=~\int d^4 x \left( i~\epsilon^{\mu \nu \alpha \beta}~
B_{\mu \nu} \ast \partial_{\alpha} A_{\beta} + B_{\mu \nu} \ast B^{\mu \nu} + 
\frac{\alpha}{2}~b \ast b - b \ast \partial_{\mu} A^{\mu} + 
\bar{c} \ast \Box c \right). 
\end{eqnarray} 
While the interaction action is given by :     
\begin{eqnarray}
S_{int} & = & \int d^4 x \left( \frac{i g}{2}~\epsilon^{\mu \nu \alpha \beta}~
B_{\mu \nu} \ast \{A_{\alpha}, A_{\beta} \} + g~
\bar{c} \ast \partial_{\mu} \{ A^{\mu}, c \} \right).
\end{eqnarray} 
Green's functions are derived by using the generating functional 
$Z_0\left[ J_{A~\mu}, J_{B~{\mu \nu}}, J_b, J_c, J_{\bar{c}} \right]$ with 
sources terms for each of the fields : 
\begin{eqnarray}
Z_0\left[ J_{A~\mu}, J_{B~{\mu \nu}}, J_b, J_c, J_{\bar{c}} \right] = 
{\cal N}~\int {\cal D} A {\cal D} B {\cal D} b {\cal D} c 
{\cal D} \bar{c}~e^{ - S_0 - S_s}.
\end{eqnarray}
where 
\begin{eqnarray}
S_s & = & \int d^4 x \left( A^{\mu} \ast J_{A~\mu} + 
B^{\mu \nu} \ast J_{B~{\mu \nu}} + b \ast J_b + c \ast J_c + 
\bar{c} \ast J_{\bar{c}} \right). 
\end{eqnarray}
is the source term action. \\
Since $S_0$ is not diagonal in fields, and cannot be diagonalized, 
the resulting propagators  
of the fields are not diagonal either. This means that there are cross--
propagators that do not vanish at tree level. To derive explicitly 
Green's functions for various fields, we work in momentum space and 
shift all the fields by field independent functions $C$ : 
\begin{eqnarray}
A_{\mu} (p) \rightarrow A_{\mu} (p) + C_{A~\mu} (p)  \nonumber \\
B_{\mu \nu} (p) \rightarrow B_{\mu \nu} (p) + C_{B~{\mu \nu}} (p)  \nonumber \\
b (p) \rightarrow b (p) + C_b (p) \nonumber \\
c (p) \rightarrow c (p) + C_c (p) \nonumber \\
\bar{c} (p) \rightarrow \bar{c} (p) + C_{\bar{c}} (p). 
\end{eqnarray} 
By making the linear terms in the fields vanish and solving for $C$'s,  
the following Feynman rules for propagators $A, B, c$ 
and $\bar{c}$ are then obtained : \\ 
\begin{center}
\begin{picture}(400,50)(10,0)
\thicklines
\Gluon(60,0)(120,0){2}{8}
\Text(50,5)[]{$\mu, p \rightarrow$}
\Text(130,5)[]{$\leftarrow q, \nu$}
\Text(280,0)[]{$ D^{AA}{_{\mu \nu}} (p,q)~=~
\frac{1}{p^2}~\left( \delta_{\mu \nu} + ( \alpha - 1)
\frac{p_{\mu} p_{\nu}}{p^2} \right)~\delta ( p + q )$ }
\end{picture}
\begin{picture}(400,50)
\Gluon(60,0)(90,0){2}{4}
\Line(90,0)(120,0)
\Text(50,5)[]{$\alpha, p \rightarrow$}
\Text(130,5)[]{$\leftarrow q, \mu \nu$}
\Text(280,0)[]{$D^{BA}{_{\mu \nu \alpha}} (p,q)~=~
- \frac{1}{2}~\epsilon_{\mu \nu \rho \alpha}~
\frac{p^{\rho}}{p^2}~\delta ( p + q )$}
\end{picture} 
\begin{picture}(400,50)
\Line(60,0)(120,0)
\Text(50,5)[]{$\mu \nu, p \rightarrow$}
\Text(130,5)[]{$\leftarrow q, \alpha \beta$}
\Text(280,0)[]{$D^{BB}{_{\mu \nu \alpha \beta}} (p,q)~=~
- \frac{1}{4}~\epsilon_{\mu \nu \lambda \gamma}~
\epsilon_{\alpha \beta \rho}{^{\gamma}}~
\frac{p^{\lambda} p^{\rho}}{p^2}~\delta( p + q)$}
\end{picture}
\begin{picture}(400,50)
\DashLine(60,0)(120,0){2}{}
\Text(50,5)[]{$p \rightarrow$}
\Text(130,5)[]{$\leftarrow q$}
\Text(280,0)[]{$D^{c \bar{c}} (p, q)~=~
- \frac{i}{p^2}~\delta ( p + q )$ .}
\end{picture} \\
~\\
~\\
{\sl \hskip 10pt {\bf figure 1}}
\end{center} 
~\\
~\\
Interaction vertices arise from $S_{int}$ as usual. Indeed, in momentum 
space $BAA$ and $c A \bar{c}$ vertices are :
\begin{center}
\begin{picture}(400,50)
\Gluon(60,0)(120,0){2}{10}
\Line(90,0)(90,30)
\Text(50,5)[]{$r~\beta \rightarrow$}
\Text(130,5)[]{$\leftarrow q~\alpha$}
\Text(70,32)[]{$\mu \nu ,p \downarrow$}
\Text(280,0)[]{$V^{BAA}{_{\mu \nu \alpha \beta}} (p,q,r)~=~
- 2 g ~\epsilon_{\mu \nu \alpha \beta}~\sin \omega (p,q)~
\delta (p + q + r)$}
\end{picture}
\begin{picture}(400,50)
\DashLine(60,0)(120,0){2}{}
\Gluon(90,0)(90,30){2}{4}
\Text(50,5)[]{$r \rightarrow$}
\Text(130,5)[]{$\leftarrow q$}
\Text(70,32)[]{$\mu,p \downarrow$}
\Text(280,0)[]{$V^{c A \bar{c}}{_{\mu}} (p,q,r)~=~
- 2 i g~ q_{\mu} \sin \omega (p,q)~ \delta ( p + q + r )$ .}
\end{picture} \\ 
~\\
~\\
{\sl \hskip 10pt {\bf figure 2}}
\end{center} 
~\\
~\\ 
The above Feynman rules for $U(1)$ BFYM on noncommutative 
${\mathbb{R}}^4$ are the same 
as the usual euclidean BFYM on commutative 
${\mathbb{R}}^4$ for $SU(N)$ Lie algebra 
in which group indices $a$ are identified with the momentum $p_a$ and 
the structure constants $f_{abc}$ with 
$- 2 i \sin \omega (p_b,p_c)~\delta ( p_a + p_b + p_c )$ ,~( see 
~\cite{doug} and ~\cite{mar,she} for Yang--Mills theory on 
noncommutative torus and noncommutative ${\mathbb{R}}^4$ ). \\ 
In figure 3 and 4 ( see Appendix ), all relevant one--loop diagrams  
for self--energies and vertices are presented.  
We have explicitly calculated the one--loop diagrams of BFYM on noncommutative 
${\mathbb{R}}^4$ where only planar contributions are considered. Every 
diagram gets then multiplied by a phase factor that depends only on the 
momenta of the external lines. We point out that non--planar contributions 
may create some trouble at higher order. \\ 
These calculations are done by using dimensional regularization in 
$D = 4 - 2 \epsilon$ dimension~\cite{vel,ber} where the 4--dimensional 
measure $\frac{d^4 k}{(2 \pi)^4}$ is replaced by the $D$--dimensional 
one $\frac{d^D k}{((2 \pi)^D}$ and before performing the 
$D$--dimensional integration, some care has to be taken according to the 
rules in~\cite{ber}. \\ 
Moreover, whenever a product of sines appears, we express each 
$\sin \omega (p,q)$ as $\frac{1}{2 i}~( e^{i \omega (p,q)} - 
e^{- i \omega (p,q)} )$ and transform these sines into a first term 
depending on the internal momentum which provides the Feynman integrals 
by oscillatory factors making them finite and a second term 
independent of the internal momentum   
so that the integrals have poles. Then we extract the one--loop  
UV divergent contribution to all the divergent $1PI$ Green functions. \\ 
The obtained results for self--energies are then given as follows : \\ 
a) $AA$ self energy,     
\begin{eqnarray*}
\tilde{D}^{AA}{_{\mu \nu}} (p, q) & = & ( \frac{1}{3} - \alpha )~ 
\frac{g^2}{(4 \pi)^2}~\Gamma ( \epsilon )~ \left( p^2 \delta_{\mu \nu} - 
p_{\mu} p_{\nu} \right)~\delta ( p + q ). 
\end{eqnarray*}
b) $A B$ self energy, 
\begin{eqnarray*}
\tilde{D}^{A B}{_{ \alpha \beta \mu}} (p ,q ) & = & \frac{1}{2}~( 3 - \alpha )~
\frac{g^2}{(4 \pi)^2}~\Gamma ( \epsilon )~\epsilon_{\alpha \beta \rho 
\mu}~p^{\rho}~\delta ( p + q ). 
\end{eqnarray*}
c) $B B$ self energy, 
\begin{eqnarray*}
\tilde{D}^{BB}{_{\mu \nu \alpha \beta}} (p ,q ) & = &  
- (1 + \alpha )~\frac{g^2}{(4 \pi)^2}~\Gamma ( \epsilon )~ 
\left( \delta_{\mu \alpha}~\delta_{\nu \beta} - 
\delta_{\mu \beta}~\delta_{\nu \alpha} \right)~\delta ( p + q ).
\end{eqnarray*}
d) ghosts $c \bar{c}$ self energy,  
\begin{eqnarray*}
\tilde{D}^{c \bar{c}} (p ,q) & = &  
\frac{3}{2}~( 1 - \alpha )~\frac{g^2}{(4 \pi)^2}~\Gamma ( \epsilon )~p^2~
\delta ( p + q). 
\end{eqnarray*}
Similarly the divergent parts for the two vertices $BAA$ and 
$\bar{c} A c$ read, \\ 
\begin{eqnarray*}
\tilde{V}^{BAA}{_{\mu \nu \alpha \beta}} (p, q, r) & = &  
2~\alpha~\frac{g^3}{( 4 \pi)^2}~\Gamma ( \epsilon)~\epsilon_{\mu \nu 
\alpha \beta}~\sin \omega (p, q)~\delta ( p + q + r ).
\end{eqnarray*}
\begin{eqnarray*}
\tilde{V}^{\bar{c} A c}{_{\mu}} (p, q, r) & = & 
4~i~\alpha~\frac{g^3}{(4 \pi)^2}~\Gamma ( \epsilon)~q_{\mu}~
\sin \omega (p, q)~\delta ( p + q + r).   
\end{eqnarray*} 
We see from the power counting argument, that the superficial degree of 
divergence $\omega_s$ for these diagrams is given by : 
\begin{eqnarray*}
\omega_s & = & 4 - ( E_A + E_c ) - 2 E_B .
\end{eqnarray*}
where $E_A$, $E_B$ and $E_c$ represent the number of $A$, $B$ and ghost 
$c$ external lines respectively. \\ 
Then we remark that the one--loop divergent contributions for self--energies 
and vertices are like the usual BFYM theory on commutative 
${\mathbb{R}}^4$ for 
the Lie algebra $SU(N)$ with structure constants $f_{abc}$ replaced by 
$- 2 i \sin \omega (p_b, p_c)$ and the quadratic Casimir $C_2( G )$ 
equal to 2. \\
Up on these replacements, our results agree with the usual $SU(N)$ BFYM 
on commutative ${\mathbb{R}}^4$~\cite{zen} where their calculations  
have been done in the Landau gauge ( $\alpha = 0$ ). \\
We see also that the divergent part of the ghost vertex 
$\tilde{V}^{\bar{c} A c} (p,q,r)$ vanishes in the Landau gauge due to the 
transversality of the propagators in this gauge and the vertex 
$\tilde{V}^{BAA} (p,q,r)$ is finite. \\
At the one--loop level, two other trilinear and quadrilinear vertices  
$\tilde{V}^{AAA}_{\mu \nu \alpha} (p_1,p_2,p_3)$ and  
$\tilde{V}^{AAAA}_{\mu \nu \alpha \beta} (p_1,p_2,p_3,p_4)$ respectively    
appear and do not belong to the tree--level $U(1)$ BFYM action on 
noncommutative ${\mathbb{R}}^4$. They correspond to the nonlinear 
self--interactions of $U(1)$ Yang--Mills action on 
noncommutative ${\mathbb{R}}^4$ and arise from term  
$\left( F^{\mu \nu} \ast F_{\mu \nu} \right) (x)$ which is allowed to 
enter at the quantum level due to the symmetries of the theory. We will 
see in the next section how such a counter--term can arise.  
\section{ Renormalization and $\beta$-- function : }
Before we perform the renormalization, we consider the gauge fixing 
action $S_{gf}$ and integrate out the auxiliary fields $b$. After this 
manipulation, the $S_{BFYM} + S_{gf}~=~S$ becomes : 
\begin{eqnarray}
S = \int d^4 x \left(~\frac{i}{2}~\epsilon^{\mu \nu \alpha \beta}~
B_{\mu \nu} \ast F_{\alpha \beta} + B_{\mu \nu}  \ast B^{\mu \nu} - 
\frac{1}{2 \alpha} (\partial_{\mu} A^{\mu} ) \ast (\partial_{\nu} A^{\nu}) + 
\bar{c} \ast \partial_{\mu} D^{\mu} c \right). 
\end{eqnarray}
To perform the renormalization, we follow standard techniques~\cite{zub,mut}  
and substitute the bare quantities for the renormalized 
ones, where in general an operational mixing is allowed by the symmetry and 
parity properties of the fields. We then have for the fields :  
\begin{eqnarray}
\left( \begin{array}{c} 
B_{0~\mu \nu} \\ 
F_{0~\mu \nu} \end{array} \right) & = & \left( \begin{array}{cc} 
Z_{BB}~\delta_{\mu \alpha} \delta_{\nu \beta} & \frac{i}{2}~
Z_{BA}~\epsilon_{\mu \nu \alpha \beta} \\
0 & Z_{AA}~\delta_{\mu \alpha} \delta_{\nu \beta} \end{array} \right)~
\left( \begin{array}{c} 
B_R{^{\alpha \beta}} \\
F_R{^{\alpha \beta}} \end{array} \right) \nonumber \\
c_0 & = & Z_c~c_R   \nonumber \\
\bar{c}_0 & = & Z_c~\bar{c}_R.
\end{eqnarray} 
and the parameters $g$ and $\alpha$ : 
\begin{eqnarray}
g_0~~=~~Z_g~g_R~,~~~\alpha_0~=~Z_{\alpha}~\alpha_R. 
\end{eqnarray}  
where the constants $Z_{AA}, Z_{BA}$ and $Z_{BB}$ and $Z_c$ are the 
$A$ field, $BA$ fields, $BB$ fields and ghost--field renormalization 
constants, respectively, while the constants 
$Z_g$ and $Z_{\alpha}~=~Z_{AA}{^2}$ are the 
coupling constant and gauge parameter renormalization constants. 
Here $F_{R~\mu \nu}~=~\partial_{\mu} A_{R~\nu} - \partial_{\nu} A_{R~\mu} 
+ Z_{AA} Z_g g_R~\{ A_{R~\mu}, A_{R~\nu} \}$ is the renormalized field 
strength. \\   
Notice that the presence of the factor $i$ 
is necessary in (21) since $B_{\mu \nu}$ and 
$F_{\mu \nu}$ have opposite parity and a mixing of $B_{R~\mu \nu}$ is 
not allowed since $F_{\mu \nu}$ must be a curvature tensor. With the 
above field redefinitions, a counter--term 
$\left( F_{\mu \nu} \ast F^{\mu \nu} \right) (x)$, absent at tree level 
in the theory, appears at the quantum level which is required to 
renormalize the trilinear $AAA$ and quadrilinear $AAAA$ vertices arising at 
one--loop level. \\
~\\
Now let us analyze the structure of the counter--terms that make finite 
the self--energies and the vertices of the $U(1)$ BF--Yang--Mills theory 
on noncommutative ${\mathbb{R}}^4$. It is necessary first to begin by 
writing the action in terms of bare quantities : 
\begin{eqnarray}
S = \int d^4 x \left( \frac{i}{2} \epsilon^{\mu \nu \alpha \beta} 
B_{0~\mu \nu} \ast F_{0~\alpha \beta} + B_{0~\mu \nu} \ast B_0{^{\mu \nu}} - 
\frac{1}{2 \alpha_0} (\partial_{\mu} A_0{^{\mu}}) \ast 
(\partial_{\nu} A_0{^{\nu}}) + 
\bar{c}_0 \ast \partial_{\mu} D^{\mu} c_0 \right). 
\end{eqnarray}
and then in terms of renormalized fields and renormalization constants : 
\begin{eqnarray}
S~=~\int~d^4 x~\left( \frac{i}{2}~Z_{BB}~( Z_{AA} + 2 Z_{BA} )~
\epsilon^{\mu \nu \alpha \beta}~B_{R~\mu \nu} \ast F_{R~\alpha \beta}~
+~ Z_{BB}{^2}~B_{R~\mu \nu} \ast B_R{^{\mu \nu}} \right. \nonumber \\
\left. -~ Z_{BA}~( Z_{AA} + Z_{BA} )~
F_{R~\mu \nu} \ast F_R{^{\mu \nu}} - \frac{1}{2 \alpha_R} 
(\partial_{\mu} A_R{^{\mu}}) \ast (\partial_{\nu} A_R{^{\nu}}) + 
Z_c{^2} \bar{c}_R \ast \partial_{\mu} D^{\mu} c_R  
~\right).
\end{eqnarray}  
where $D_{\mu} c_R~=~\partial_{\mu} c_R + Z_{AA} Z_g g_R 
\{ A_{R~\mu}, c_R \}$ is the renormalized covariant derivative. \\
Apart from the constraints on the renormalization constants by the gauge 
Ward identities, the $Z$'s are in principle completely arbitrary. In 
practice, because the $U(1)$ BFYM theory on noncommutative 
${\mathbb{R}}^4$ needs regularizing, the arbitrariness of the $Z$'s is 
only in the finite parts. Since Feynman rules of the $U(1)$ BFYM theory 
on noncommutative ${\mathbb{R}}^4$ at tree level should not be modified 
we expect : 
\begin{eqnarray}
Z_{AA} & \simeq & 1 + a ( \alpha_R)~\frac{g_R{^2}}{(4 \pi)^2}~
\Gamma (\epsilon )~+~ ( \cdots ) ~+~ O ( g_R{^2} ) \nonumber \\
Z_{BB} & \simeq & 1 + b ( \alpha_R)~\frac{g_R{^2}}{(4 \pi)^2}~
\Gamma (\epsilon )~+~ ( \cdots ) ~+~ O ( g_R{^2} ) \nonumber \\ 
Z_{BA} & \simeq & c ( \alpha_R)~\frac{g_R{^2}}{(4 \pi)^2}~
\Gamma (\epsilon )~+~ ( \cdots ) ~+~ O ( g_R{^2} )  \nonumber \\
Z_c & \simeq & 1 + d ( \alpha_R)~\frac{g_R{^2}}{(4 \pi)^2}~
\Gamma (\epsilon )~+~ ( \cdots ) ~+~ O ( g_R{^2} ).
\end{eqnarray}
where $(\cdots )$ represents finite terms at order $g_R{^2}$.  
Note that $Z_{BA} \sim g_R{^2}~\Gamma (\epsilon)$ in order not to modify     
the Feynman rules at the tree level. Moreover, the $Z$'s renormalization 
constants should be determined by adjusting the counter--terms so as    
to cancel overall divergences appearing in one--loop Feynman 
amplitudes. In our case, they are obtained by straightforward 
comparison between the Feynman rules for the quadratic (24) and the 
divergent parts of self--energies. \\
Consequently, the following system appears : 
\begin{eqnarray}
Z_{BB}~( Z_{AA} + 2 Z_{BA} ) & = & 1 + \frac{1}{2} ( 3 - \alpha_R)~
\frac{g_R{^2}}{(4 \pi)^2}~\Gamma (\epsilon )~+~
( \cdots ) ~+~ O ( g_R{^2} ) \nonumber \\ 
Z_{BB}{^2} & = & 1 - ( 1 - \alpha_R )~\frac{g_R{^2}}{(4 \pi)^2}~
\Gamma (\epsilon )~+~( \cdots ) ~+~ O ( g_R{^2} ) \nonumber \\
4 Z_{BA}~( Z_{AA} + Z_{BA} ) & = & ( \frac{1}{3} - \alpha_R)~
\frac{g_R{^2}}{(4 \pi)^2}~\Gamma (\epsilon )~+~
( \cdots ) ~+~ O ( g_R{^2} ) \nonumber \\
Z_c{^2} & = & 1 + \frac{3}{2}~( 1 - \alpha_R)~
\frac{g_R{^2}}{(4 \pi)^2}~\Gamma (\epsilon )~+~
( \cdots ) ~+~ O ( g_R{^2} ).
\end{eqnarray} 
Solving (26) by a use of (25) gives :
\begin{eqnarray}
a (\alpha_R) & = & \frac{1}{2}~( \frac{13}{3} - \alpha_R )~,~~
b (\alpha_R)~=~-~\frac{1}{2} (1 + \alpha_R )~,~~ 
c (\alpha_R)~=~-~\frac{1}{4} (\frac{1}{3} - \alpha_R )~,~ \nonumber \\
d (\alpha_R) & = & \frac{3}{4}~( 1 - \alpha_R ).
\end{eqnarray} 
The structure of renormalization constants that remove the divergent 
quantities are : 
\begin{eqnarray}
Z_{AA} & \simeq & 1 + \frac{1}{2}~( \frac{13}{3} - 
\alpha_R )~\frac{g_R{^2}}{(4 \pi)^2}~\Gamma (\epsilon ) \nonumber \\
Z_{BB} & \simeq & 1 - ~\frac{1}{2}~( 1 +  
\alpha_R )~\frac{g_R{^2}}{(4 \pi)^2}~\Gamma (\epsilon ) \nonumber \\ 
Z_{BA} & \simeq & - ~\frac{1}{2}~( \frac{1}{3} -
\alpha_R )~\frac{g_R{^2}}{(4 \pi)^2}~\Gamma (\epsilon ) \nonumber \\ 
Z_c & \simeq & 1 + \frac{3}{4}~( 1 -
\alpha_R )~\frac{g_R{^2}}{(4 \pi)^2}~\Gamma (\epsilon ). 
\end{eqnarray}
Our remaining task is now to determine the renormalization constant $Z_g$.  
Indeed from the $c A \bar{c}$ vertex we get :
\begin{eqnarray}
Z_g~Z_{AA}~Z_c{^2} & = & 1 - 2~\alpha_R~ 
\frac{g_R{^2}}{(4 \pi)^2}~\Gamma (\epsilon ). 
\end{eqnarray} 
It is easy to extract from (29) the renormalization of the coupling constant   
which turns out to be :
\begin{eqnarray}
Z_g & = & 1 -~\frac{11}{3}~\frac{g_R{^2}}{(4 \pi)^2}~\Gamma (\epsilon ). 
\end{eqnarray} 
The $\beta$--function can now be easily read from (30) : 
\begin{eqnarray}
\beta_1 & = & -~\frac{11}{3}.
\end{eqnarray}
which ensures that the theory is asymptotically free~\cite{mut}. 
Moreover, as expected, the UV--behaviour of the $U(1)$ BFYM on 
noncommutative ${\mathbb{R}}^4$ is the same as the usual $SU(N)$ BFYM 
and Yang--Mills theories on commutative ${\mathbb{R}}^4$. \\
We also notice that the Weyl--Moyal matrix $\theta_{\mu \nu}$ expressing 
the non--local caracter of the interaction is not renormalized at the 
one--loop order.
\section{Conclusions :} 
In summary we have introduced the $U(1)$ BFYM theory on noncommutative 
${\mathbb{R}}^4$ and shown its equivalence to $U(1)$ Yang--Mills  on 
noncommutative ${\mathbb{R}}^4$ after integrating out the 
antisymmetric $B_{\mu \nu}$ field. \\
Quantization of this theory in the BRST symmetry formalism  is studied 
where the full quantum action with the Slavnov--Taylor identity is 
obtained. \\
After extracting the Feynman rules, one--loop calculations have been 
performed and particularly the one--loop UV--divergent contribution 
to the divergent $1 PI$ Green functions. Its renormalization at one--loop 
level has been described and its asymptotical free behaviour checked. 
Moreover we have shown that the UV--behaviour of the $U(1)$ BFYM theory 
on noncommutative ${\mathbb{R}}^4$ is similar to the usual $SU(N)$ 
BFYM and Yang--Mills theories on commutative ${\mathbb{R}}^4$. We have also 
seen that -- at least -- at the one--loop order no renormalization of the 
Weyl--Moyal matrix $\theta_{\mu \nu}$ is needed. \\
Within the one--loop level, apart from some complications in the 
computations due to the non--local terms in the action that result in     
Feynman integrals with oscillatory factors, no new phenomena appear. 
It becomes clear that the $U(1)$ BFYM on noncommutative 
${\mathbb{R}}^4$ behaves like the commutative case. \\
~\\
For $g = 0$, the action (5) is just the $U(1)$ BF theory on noncommutative 
${\mathbb{R}}^4$~\footnote{ For commutative spaces, the BF is a 
topological theory of Schwarz type. Indeed for this theory on $S^4$ or 
${\mathbb{R}}^4$, the partition function is just one whereas for $T^4$, 
it is proportional to $vol( T^{\ast} )$ where $T^{\ast}$ is the dual torus. 
A question -- that can be adressed - is the value of the partition 
function for different topologies on noncommutative spaces. Thanks to 
Sheikh--Jabbari for giving rise to this point. }
: 
\begin{eqnarray}
S_{BF} & = & \int d^4 x~\frac{i}{2}~\epsilon^{\mu \nu \alpha \beta}~
B_{\mu \nu} \ast F_{\alpha \beta}. 
\end{eqnarray}
As for the commutative BF, this $U(1)$ BF on noncommutative ${\mathbb{R}}^4$ 
action is invariant under the gauge symmetry : 
\begin{eqnarray} 
\delta_{\epsilon} A_{\mu}~=~D_{\mu} \epsilon~,~
\delta_{\epsilon} F_{\mu \nu}~=~\{ F_{\mu \nu}, \epsilon \}~,~
\delta_{\epsilon} B_{\mu \nu}~=~\{ B_{\mu \nu}, \epsilon \}. 
\end{eqnarray}
and an extra ''topological '' symmetry : 
\begin{eqnarray}
\delta_{\psi} A_{\mu}~=~0~,~
\delta_{\psi} F_{\mu \nu}~=~0~,~
\delta_{\psi} B_{\mu \nu}~=~D_{\mu} \psi_{\nu} - D_{\nu} \psi_{\mu}. 
\end{eqnarray}
Among the problems -- that could be studied-- are the quantization and 
the perturbative renormalization of the BF theory on noncommutative spaces. 
These will be adressed elsewhere. \\ 
We should mention that the $U(1)$ BFYM on noncommutative ${\mathbb{R}}^4$ 
can also be interpreted as a deformation ( perturbation ) of the pure 
$U(1)$ BF theory on noncommutative ${\mathbb{R}}^4$ 
having besides the gauge symmetry, the topological one. \\ 
Prototype of topological gauge field theory of Schwarz type is the 
Chern--Simons theory. Renormalization and finiteness of this system on 
noncommutative $3d$ spaces can also be carried out and will be 
reported in a future work. \\
~\\ 
In ref~\cite{sei}, Seiberg and Witten established a relation between 
noncommutative Yang--Mills and commutative one. Indeed, they obtained a 
transformation from commutative gauge field $A$ with gauge parameter 
$\lambda$ to noncommutative gauge field $\hat{A}$ with gauge parameter 
$\hat{\lambda}$ by requiring the equivalence of the gauge transformation 
of $A$ and $\hat{A}$. We have checked and found that this mapping exists also 
for BF--Yang Mills theory~\cite{ben}. 
\section*{Acknowledgment} 
I would like to thank the DAAD for its financial support and 
A. Davydychev for the axodraw file. I'm grateful to  
R. Coquereaux, T. Krajewski, C.P. Martin, F. Scheck and M.M. Sheikh--Jabbari 
for reading the manuscript and their comments. I thank also I. Chepelev 
for his email concerning the new version his work~\cite{che}, 
J.M. Gracia--Bondia for discusions and the referee 
for his remarks. 
\newpage
\section*{Appendix :}
Here are the different one loop Feynman diagrams relevant for BFYM theory : 
~\\
\begin{center}
\begin{picture}(400,50)(10,0)
\thicklines
\Gluon(10,0)(30,0){2}{3}
\GCirc(40,0){10}{0}
\Gluon(50,0)(70,0){2}{3}
\Text(15,10)[]{$A$}
\Text(65,10)[]{$A$}
\Text(80,0)[]{ = }
\Gluon(100,0)(120,0){2}{3}
\CArc(140,0)(20,90,180)
\GlueArc(140,0)(20,0,90){2}{5}
\GlueArc(140,0)(20,180,270){2}{5}
\CArc(140,0)(20,270,360)
\Gluon(160,0)(180,0){2}{3}
\Text(190,0)[]{+  }
\Gluon(200,0)(220,0){2}{3}
\DashArrowArc(235,0)(15,180,0){3}
\DashArrowArc(235,0)(15,0,180){3}
\Gluon(250,0)(270,0){2}{3}
\Text(280,0)[]{ + }
\Gluon(290,0)(310,0){2}{3}
\Line(310,0)(340,0)
\GlueArc(325,0)(15,0,180){2}{5}
\Gluon(340,0)(360,0){2}{3}
\end{picture}
~\\
~\\
\begin{picture}(400,50)
\thicklines
\Gluon(10,0)(30,0){2}{3}
\GCirc(40,0){10}{0}
\Line(50,0)(70,0)
\Text(15,10)[]{$A$}
\Text(65,10)[]{$B$}
\Text(80,0)[]{ = }
\Gluon(100,0)(120,0){2}{3}
\GlueArc(140,0)(20,0,180){2}{7}
\CArc(140,0)(20,180,270)
\GlueArc(140,0)(20,270,360){2}{5}
\Line(160,0)(180,0)
\end{picture}
~\\
~\\
\begin{picture}(400,50)
\thicklines
\Line(10,0)(30,0)
\GCirc(40,0){10}{0}
\Line(50,0)(70,0)
\Text(15,10)[]{$B$}
\Text(65,10)[]{$B$}
\Text(80,0)[]{ = }
\Line(100,0)(120,0)
\GlueArc(140,0)(20,0,180){2}{7}
\GlueArc(140,0)(20,180,360){2}{7}
\Line(160,0)(180,0)
\end{picture}
~\\
~\\
\begin{picture}(400,50)
\thicklines
\DashLine(10,0)(30,0){3}
\GCirc(40,0){10}{0}
\DashLine(50,0)(70,0){3}
\Text(15,10)[]{$c$}
\Text(65,10)[]{$\bar{c}$}
\Text(80,0)[]{ = }
\DashLine(100,0)(120,0){3}
\DashArrowLine(120,0)(160,0){3}
\GlueArc(140,0)(20,0,180){2}{7}
\DashLine(160,0)(180,0){3}
\end{picture} \\
~\\
~\\
{\sl \hskip 10 pt {\bf figure 3 : Self--energies }}
\end{center}
\begin{center}
\begin{picture}(400,50)
\thicklines
\GCirc(20,0){20}{0}
\Gluon(20,20)(20,40){2}{3}
\DashLine(2.68,-10)(-14.64,-20){3}
\DashLine(37.32,-10)(54.64,-20){3}
\Text(10,30)[]{$A$}
\Text(-20,-15)[]{$c$}
\Text(60,-15)[]{$\bar{c}$}
\Text(80,0)[]{ = }
\Gluon(120,20)(120,40){2}{3}
\DashLine(102.68,-10)(85.36,-20){3}
\DashLine(137.32,-10)(154.64,-20){3}
\DashArrowLine(102.68,-10)(137.32,-10){3}
\Gluon(120,20)(137.32,-10){2}{5}
\Line(120,20)(111.34,5)
\Gluon(111.34,5)(102.68,-10){2}{3}
\Text(160,0)[]{ + }
\Gluon(200,20)(200,40){2}{3}
\DashLine(182.68,-10)(165.36,-20){3}
\DashLine(217.32,-10)(234.64,-20){3}
\Gluon(182.68,-10)(217.32,-10){2}{5}
\DashArrowLine(200,20)(217.32,-10){3}
\DashArrowLine(182.68,-10)(200,20){3}
\Text(300,0)[]{$+~~~~permutations$}
\end{picture}
\end{center}
~\\
\begin{center}
\begin{picture}(400,50)
\thicklines
\GCirc(20,0){20}{0}
\Line(20,20)(20,40)
\Gluon(2.68,-10)(-14.64,-20){2}{3}
\Gluon(37.32,-10)(54.64,-20){2}{3}
\Text(10,30)[]{$B$}
\Text(-20,-15)[]{$A$}
\Text(60,-15)[]{$A$}
\Text(80,0)[]{ = }
\Line(120,20)(120,40)
\Gluon(102.68,-10)(85.36,-20){2}{3}
\Gluon(137.32,-10)(154.64,-20){2}{3}
\Line(102.68,-10)(137.32,-10)
\Gluon(120,20)(137.32,-10){2}{5}
\Gluon(120,20)(102.68,-10){2}{5}
\Text(160,0)[]{ + }
\Line(200,20)(200,40)
\Gluon(182.68,-10)(165.36,-20){2}{3}
\Gluon(217.32,-10)(234.64,-20){2}{3}
\Line(182.68,-10)(200,-10)
\Gluon(200,-10)(217.32,-10){2}{3}
\Gluon(200,20)(208.66,5){2}{3}
\Line(208.66,5)(217.32,-10)
\Gluon(200,20)(182.68,-10){2}{5}
\Text(300,0)[]{$+~~~~permutations$}
\end{picture} \\
~\\
~\\
{\sl \hskip 10 pt {\bf figure 4 : Vertex} }
\end{center}

\end{document}